\documentclass[floats,twocolumn,prb]{revtex4}
\usepackage{graphicx}
\begin{document}
\title{Variational Monte Carlo Study of Anderson Localization in the Hubbard Model}
\author{A. Farhoodfar$^{1,2}$}
\author{R. J. Gooding$^{1}$}
\author{W. A. Atkinson$^2$}\email{billatkinson@trentu.ca}
\affiliation{$^1$Department of Physics and Astronomy, Queen's University, Kingston, Ontario, Canada, K7L 3N6 }
\affiliation{$^2$Department of Physics and Astronomy, Trent University, Peterborough, Ontario, Canada K9J 7B8}
\date{\today}
\begin{abstract}
We have studied the effects of interactions on persistent currents in half-filled and quarter-filled Hubbard models with weak and intermediate strength disorder.  Calculations are performed using a variational Gutzwiller ansatz that describes short range correlations near the Mott transition.  We apply an Aharonov-Bohm magnetic flux, which generates a persistent current that can be related to the Thouless conductance.  The magnitude of the current depends on both the strength of the screened disorder potential and the strength of electron-electron correlations, and the Anderson localization length can be extracted from the scaling of the current with system size.  At half filling, the persistent current is reduced by strong correlations  when the interaction strength is large.  Surprisingly, we find that the disorder potential is strongly  screened in the large interaction limit, so  that the localization length grows with increasing interaction strength even as the magnitude of the current is suppressed.  This supports earlier dynamical mean field theory predictions that the elastic scattering rate is suppressed near the Mott transition.
\end{abstract}
\maketitle

\section{Introduction}

The Hubbard model is the standard model for strongly-correlated particles in a tight-binding lattice, and has been widely useful as a minimal model to describe  a variety of transition metal oxides,\cite{Imada1998} quantum dot arrays\cite{Stafford1994} and,  more recently, trapped atomic gases.\cite{Lewenstein2007}  Real materials are disordered and extensions of the Hubbard model, most notably the Anderson-Hubbard model (AHM), have been introduced to study the subtle interplay between disorder and interactions.  Like the Hubbard model, the AHM has not been solved exactly, except in special cases, and appears to have a rich phase diagram.

Over the past decade, there has been considerable interest in Anderson localization in strongly-correlated metals near half filling.  At half filling,  Coulomb interactions can induce a gapped ``Mott insulating" state, and the question is how states near the Mott state are affected by strong interactions.   This is particularly relevant to the cuprate high temperature superconductors, where the electronic properties are tuned by chemical doping.  Chemical doping, in addition to changing the carrier density (and therefore proximity to the Mott state), disorders the materials.    Experiments  have generically found states exhibiting characteristics of Anderson localization that lie between the metallic and Mott insulating phases.\cite{Segawa1999,Lupi2009,Sun2009}   The location of the metal insulator transition (MIT) appears to be a function of both the itinerant hole density and the level of disorder, but not the residual ($T\rightarrow 0$) resistivity,\cite{Segawa1999} which hints at a possible failure of the conventional one-parameter scaling theory.\cite{Lu2008}   More generally, these experiments raise the possibility that the Anderson MIT in strongly correlated materials may differ qualitatively from that in conventional weakly correlated metals.

In this work, we use the AHM to study the effects of strong correlations on Anderson localization. The Anderson-Hubbard Hamiltonian is
\begin{equation}
\hat H = \sum_{\langle i,j\rangle,\sigma} t_{ij} c^\dagger_{i\sigma}c_{j\sigma} + \sum_i \left (U\hat n_{i\uparrow} \hat n_{i\downarrow} + V_i \hat n_i \right ),
\label{eq:ham}
\end{equation}
where the indices $i$ and $j$ refer to sites on a $d$-dimensional lattice,
where $c_{i\sigma}$ annihilates a spin-$\sigma$ electron from site $i$, $\hat n_{i\sigma} = c^\dagger_{i\sigma}c_{i\sigma}$ is the number operator for site $i$, and where $\hat n_i = \sum_\sigma \hat n_{i\sigma}$.   We restrict ourselves to nearest-neighbor hopping, so that $t_{ij} = -t$ if $i$ and $j$ are nearest neighbors and $t_{ij}=0$ otherwise.   Disorder is introduced via the site energies $V_i$, which are chosen randomly from a uniform distribution of width $W$ (so $-W/2 \leq V_i \leq W/2$).  The AHM is therefore characterized by three energy scales: the kinetic energy $t$, the short range Coulomb interaction $U$, and the disorder strength $W$.   

\begin{figure}
\includegraphics[width=\columnwidth]{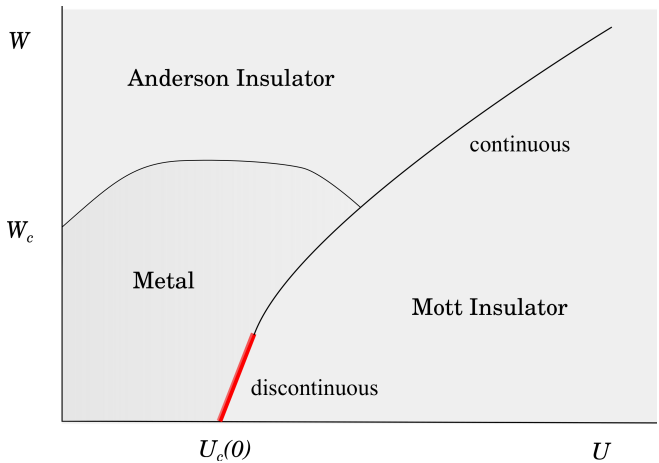}
\caption{Schematic phase diagram for the Anderson-Hubbard model in three dimensions at half filling.  Thin lines indicate continuous transitions between metallic, Anderson insulating and Mott insulating phases;  the thick solid line indicates a region of the phase boundary where DMFT predicts a discontinuity in the density of states.}
\label{fig:phasediag}
\end{figure}

A rough schematic  phase diagram for the half-filled AHM in three dimensions is shown in Fig.~\ref{fig:phasediag} (the phase diagram in lower dimensions has not, to our knowledge, been established).  This phase diagram is based on a number of published calculations\cite{Tusch1993,Byczuk2005,Henseler2009,Semmler2011b} which, while they disagree on details, agree on general features.   Note that we have chosen to show the simplest nonmagnetic phase diagram (in which magnetic phases are suppressed) and have omitted phases that are not universally seen. 

There appears to be consensus on the strong disorder limit of the phase diagram:  as a function of increasing $U$, there is a continuous transition from a gapless Anderson insulator to a gapped Mott insulating state at $U=U_c(W)\sim W$.\cite{Tusch1993,Laad2001,Byczuk2005,Song2008,Henseler2009,Semmler2011b}   An interesting, recently-discovered wrinkle in this picture is the existence of a kinetic energy-driven zero bias anomaly in the density of states,\cite{Chiesa2008,Song2009,Shinaoka2009,Hongyi2010,Hongyi2011} which was  found in the one and two dimensional AHM and should presumably be present in higher dimensions.

There is less consensus on the phase diagram at weak and intermediate disorder.   
Some calculations have predicted that the density of states vanishes continuously at the MIT in two\cite{Henseler2009} and  three\cite{Otsuka2000,Henseler2009} dimensions while dynamical mean field theory\cite{Byczuk2005,Aguiar2009,Balzer2009} (DMFT) finds that the density of states is discontinuous across the MIT, and variational Monte Carlo\cite{Pezzoli2009,Pezzoli2010} calculations predict a first order Mott transition.  Although experiments are often complicated by the occurrence of broken-symmetry phases, paramagnetic Mott transitions in real materials are generally found to be consistent with DMFT.\cite{Kotliar2006}  The details of the $T=0$ MIT are important since they affect the electronic properties of the metal adjacent to the Mott transition. 

To illustrate this point, we consider a simple weak-coupling expression for the screened random potential $\tilde V_i$.  Within mean-field theory, $\tilde V_i = V_i + \frac 12 U(n_i - n_0)$, where
$n_0$ is the spatially-averaged charge density and $n_i = \langle \hat n_i\rangle$.  For weak
disorder, $n_i = n_0 + \sum_j \tilde V_j \partial n_i/\partial \tilde V_j$.    A rough estimate
for the charge susceptibility is $\partial n_i/\partial \tilde V_j \sim -2\rho_0\delta_{i,j}$, where $\rho_0$ is
the single-spin density of states at the Fermi energy.  Then
\begin{equation}
\tilde V_i \sim \frac{V_i}{1+\rho_0 U}. 
\label{eq:Vi}
\end{equation}
In a weak-coupling metal, $\rho_0$ depends only weakly on $U$ and the overall effect of increasing $U$ is to screen the disorder potential.   On the other hand, in a gapped insulator,
$\rho_0=0$ and the impurity potential is unscreened.   While Eq.~(\ref{eq:Vi}) applies to
weakly-correlated insulators, we also expect screening to vanish in (strongly-correlated) Mott insulators because the charge is unable to rearrange itself in response to the disorder potential.  
 The screened potential should therefore be a nonmonotonic function of $U$, obtaining a maximum value somewhere between small $U$ (where Eq.~(\ref{eq:Vi}) holds) and the critical value $U_c(W)$.   Indeed, several calculations have found that the impurity potential becomes unscreened as $U\rightarrow U_c$ for strong disorder.\cite{Henseler2008,Song2008,Henseler2009}  
 
On the other hand, DMFT predicts that, for weak disorder, a narrow quasiparticle band develops at the Fermi energy  near $U_c$, such that the density of states remains finite up to $U_c$.    It has been shown that because this  band tends to be pinned to the Fermi energy, impurity scattering within the band is reduced.\cite{Tanaskovic2003,Andrade2011}  In fact, for a particle-hole symmetric band,
%
a weak impurity potential is predicted to be {\em perfectly} screened as $U\rightarrow U_c$.
This remarkable result has, to our knowledge, only been found in DMFT calculations and has not been confirmed by other theoretical techniques.

In this work, we study persistent currents in an Anderson-Hubbard lattice threaded by an Aharonov-Bohm flux and focus on the region of strong correlations near the Mott transition.  These currents are affected by both the disorder potential (which tends to localize the quasiparticles), and by strong correlations (which reduce the quasiparticle spectral weight).  One of the main objectives of our analysis is to separate these two competing effects, and we find the paradoxical result that the localization length may grow with increasing $U$, even as the magnitude of the current is suppressed.

In Sec.~\ref{Sec:calculations}, we describe how we calculate the persistent current for a ring or torus threaded by an Aharonov-Bohm flux.  In Sec.~\ref{Sec:results} we compare clean and disordered  systems with and without strong correlations, from which we develop a qualitative sense of the relative importance of disorder screening and strong correlations.  We then show how these affect the localization length, which is extracted from the scaling of the persistent current with system size.    We present results for two cases: quarter filling where correlations are relatively weak, and half filling where correlations are strongest.  We find that, while the quarter-filling results are consistent with one-parameter scaling, there is evidence for a breakdown of one-parameter scaling at half-filling when the interaction strength $U$ is greater than the disorder strength $W$. These results are discussed in Sec.~\ref{Sec:conclusions}.

\section{Calculations}
\label{Sec:calculations}
The Hamiltonian is given by Eq.~(\ref{eq:ham}).  An Aharonov-Bohm flux $\Phi$ is introduced
through a complex phase in the hopping matrix elements \begin{equation}
t_{ij} = -t \exp\left(\frac{i2\pi\alpha x_{ij} }{L}\right),
\end{equation}
where $0<\alpha<1$, $L$ is the circumference of the ring (in one dimension) or torus (in two dimensions), 
and  $x_{ij}=\pm 1$ is the electron displacement in the $x$ direction during the hop from site $j$ to site $i$.
The parameter $\alpha$ is the magnitude of the Aharonov-Bohm flux in units of the flux quantum $\Phi_0 =hc/e$.  The magnitude $t$ of the hopping matrix element between nearest-neighbor sites  is taken to be the unit of energy throughout this work (ie.\ $t=1$).  

The variational ground state wavefunction has the Gutzwiller form
\begin{equation}
|\Psi_\mathrm{GWF}\rangle = P_G |\psi_\mathrm{ps}\rangle
\end{equation}
where $|\psi_\mathrm{ps}\rangle$ is a Slater product state wavefunction and 
$P_G = \prod_i [1-(1-g)\hat n_{i\uparrow}\hat n_{i\downarrow}]$ is the Gutzwiller projection operator.  The parameter $g$ is a variational parameter that introduces correlations into the wavefunction; the limit $g = 0$ corresponds to a state with no double occupancies, while $g=1$ gives an uncorrelated wavefunction $|\Psi_\mathrm{GWF} \rangle = |\psi_\mathrm{ps}\rangle$.  
This simple Gutzwiller wavefunction cannot, without additional Jastrow factors, describe the Mott transition;\cite{Capello2005,Yokoyama2006} nonetheless, the Gutzwiller projection is useful as a qualitative tool for understanding strong correlation physics near the Mott transition.

It is usual to take $|\psi_\mathrm{ps}\rangle$ to be the many-body ground state of some effectively noninteracting Hamiltonian, and we consider here two cases:  (i) $|\psi_\mathrm{ps}\rangle$ is the self-consistent paramagnetic Hartree-Fock ground state of $\hat H$, and (ii) $|\psi_\mathrm{ps}\rangle$ is the ground state of a noninteracting Hamiltonian with a screened disorder potential,
\begin{equation}
\hat H_\epsilon = - \sum_{\langle i,j\rangle,\sigma} t_{ij} c^\dagger_{i\sigma}c_{j\sigma} + \sum_i \frac{V_i}{\epsilon} \hat n_i.
\end{equation}
The parameter $\epsilon$ is treated as a variational parameter, on the same footing as $g$.  We call the projected wavefunctions obtained from (i) and (ii) the paramagnetic Gutzwiller wavefunction (PMGW) and the disordered filled sea Gutzwiller wavefunction (DFSGW) respectively.  

While it is possible to include more variational parameters, for example by taking $g$ and $\epsilon$ to be site dependent,\cite{Pezzoli2009,Pezzoli2010} we found previously\cite{Farhoodfar2009} that it is sufficient to include spatial inhomogeneity only in $|\psi_\mathrm{ps}\rangle$ when the disorder is not too strong; it is only for strong disorder (relative to the bandwidth) that a spatially varying $g$ and $\epsilon$ are essential.  This represents a large computational savings.  

\begin{figure}
\includegraphics[width=\columnwidth]{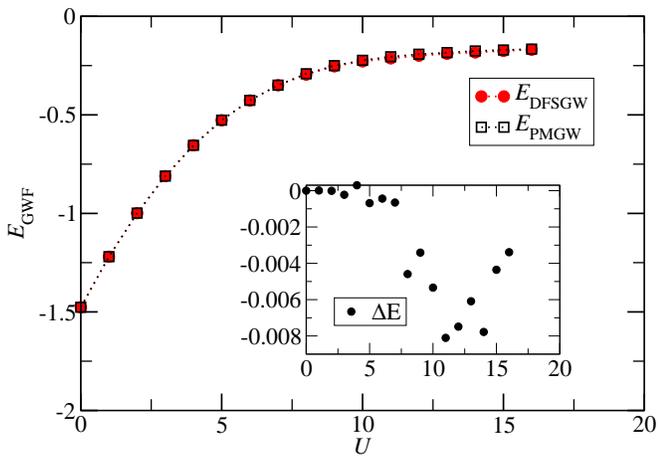}
\caption{Variational energies of the PMGW and DFSGW states for a one dimensional chain with $L=50$ and $W=4t$.  Results are at half filling, and are for a single configuration of disorder.  The energy per electron is shown.   {\em Inset:} Energy difference
$\Delta E = E_\mathrm{DFSGW}-E_\mathrm{PMGW}$ between the two variational states.}
\label{fig:energies}
\end{figure}

The  parameters $g$ and $\epsilon$ are determined by minimizing the 
energy functional
\begin{equation}
{\cal E} = \frac{\langle \Psi_\mathrm{GWF} | \hat H |\Psi_\mathrm{GWF}\rangle}{\langle \Psi_\mathrm{GWF}|\Psi_\mathrm{GWF}\rangle}.
\end{equation}
For large systems, $\cal E$ cannot be evaluated exactly, and a variational Monte Carlo (VMC) method
is used.  The idea is to expand $|\psi_\mathrm{ps}\rangle$ in terms of Fock states $|n\rangle$ (states with a definite number of electrons on each site), such that $|\psi_\mathrm{ps}\rangle = \sum_n \alpha_n |n\rangle$.  It is then simple to apply the Gutzwiller projection:  $P_G|\psi_\mathrm{ps}\rangle
= \sum_n \tilde \alpha_n|n\rangle$ where $\tilde \alpha_m = g^{D_m} \alpha_m$, with $D_m$ the number of doubly occupied sites in the ket $|m\rangle$. 
Then the variational energy can be written
\[ 
{\cal E} = \frac{ \sum_{m} |\tilde \alpha_m|^2 \sum_n \left(  H_{mn} \tilde \alpha_n/\tilde \alpha_m \right )}{\sum_p |\tilde \alpha_p|^2}.
\]  
The double sum in the numerator is computationally prohibitive to calculate and, in the simplest  approach, the sums over $m$ and $p$ are performed approximately using the Metropolis algorithm with weighting factors $|\alpha_m|^2$ and $|\alpha_p|^2$ respectively.   More sophisticated VMC algorithms exist, and in this work, we used the modified Metropolis algorithm described in Ref.~\onlinecite{Koch1999}.

The results of the energy minimization are shown in Fig.~\ref{fig:energies}, and suggest that there is little difference between the PMGW and DFSGW states. We shall see in the next section that although these two product states produce almost identical results for small $L$, they make quantitatively different (although qualitatively similar) predictions for the persistent current when $L$ is large. 
This difference stems from how screening is handled:  in the PMGW approximation, the self-consistent charge densities are obtained from a Hartree-Fock calculation {\em before} ${\cal E}$ is minimized with respect to $g$;  in the DFSGW approximation, $\cal E$ is minimized with respect to the projection $g$ and screening $\epsilon$ simultaneously.  We expect, therefore, that the PMGW state provides an upper bound on the screening, and where we have been able to compare with published results the DFSGW gives a more accurate estimate of the persistent current.   
 \begin{figure}
\includegraphics[width=\columnwidth]{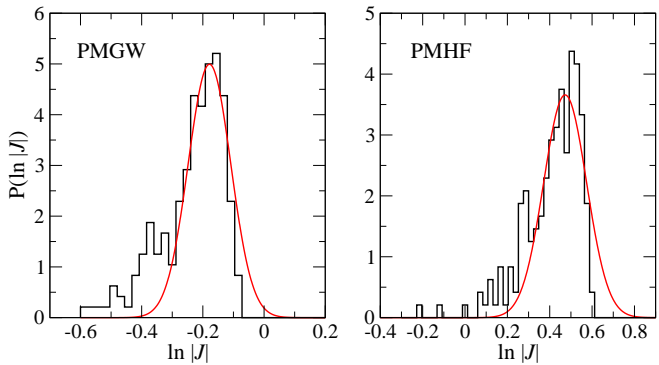}
\caption{Distribution of the persistent currents for an ensemble of one dimensional chains with $L=50$, $W=4t$ and $U=8t$.  Here and throughout this work, the Aharanov-Bohm flux is $\alpha=0.25$.  Results are shown for paramagnetic Gutzwiller (left) and paramagnetic Hartree Fock (right) approximations.  Smooth curves are Gaussian fits to the data for comparison.}
\label{fig:currentdistribution}
\end{figure}

The Aharonov-Bohm flux in the Hamiltonian generates a persistent current
\begin{equation}
J = \frac{\langle \Psi_\mathrm{GWF} | \hat J |\Psi_\mathrm{GWF}\rangle}{\langle \Psi_\mathrm{GWF}|\Psi_\mathrm{GWF}\rangle},
\end{equation}
with 
\begin{equation}
\hat J =  \frac{itea}{\hbar}\sum_{\langle i,j\rangle,  \sigma }x_{ij} \exp\left(\frac{i2\pi\alpha x_{ij} }{L}\right) c^\dagger_{i\sigma}c_{j\sigma}.
\end{equation}
We note that the persistent current is approximately related to the Thouless conductance via $g_\mathrm{Th} \sim J/\alpha$.

The persistent current depends on the length $L$ of the chain or torus, on the interaction strength, and on disorder.   In the next section, much of the discussion involves a comparison of $J$ for systems with and without disorder.  However, for one dimensional rings, it is also possible to perform a finite size scaling analysis from which the localization length can be extracted.
For localized electrons, the persistent current satisfies $J = J_0 \exp(-L/\xi)$ in the limit $L \gg \xi$, where $\xi$ is the localization length, and it is thus possible to extract the localization length from logarithmic plots of the persistent current as a function of system size.

The current $J$ depends on the details of the disorder configuration and,  in the noninteracting case, the distribution is log-normal when the system size is much larger than the localization length.\cite{Beenakker1997}  Examples of the distribution of the current values are given in Fig.~\ref{fig:currentdistribution}.  Because of the approximate log-normal current distribution, we calculate the localization length from the {\em typical} value of the current
\begin{equation}
J_\mathrm{typ} = \exp\left ( \overline {\ln J} \right )
\end{equation}
where the bar denotes an average over disorder configurations.

\section{Results}
\label{Sec:results}

\subsection{The Persistent Current}
\begin{figure}
\includegraphics[width=\columnwidth]{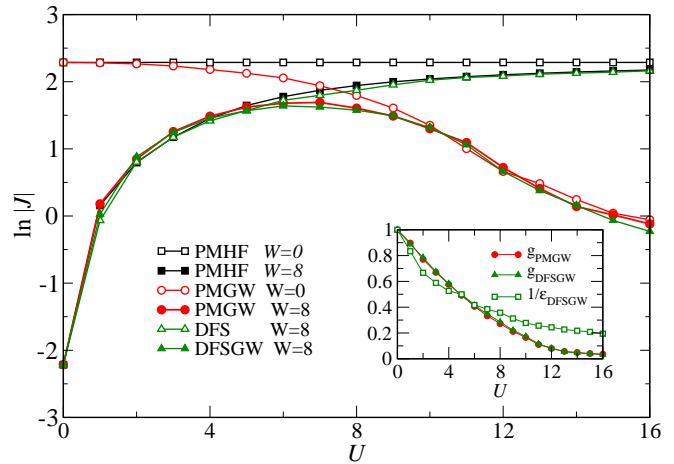}
\caption{Persistent current as a function of $U$ for  PMGW and DFSGW ground states for both a clean ($W=0$) and disordered ($W=8t$) two dimensional plane.  Results are shown at half filling ($n=1$) for a single disorder configuration with linear dimension $L=8$.    Results with $g$ set to 1 by hand include screening but not strong correlations;  these are labelled PMHF (paramagnetic Hartree-Fock, obtained from PMGW)  and DFS (disordered Fermi sea, obtained from DFSGW).  The DFS curves have the same value of $\epsilon$ as the DFSGW curves.  The inset shows the dependence of the variational parameters on $U$. }
\label{fig:J_vs_U2D}
\end{figure}

\begin{figure}
\includegraphics[width=\columnwidth]{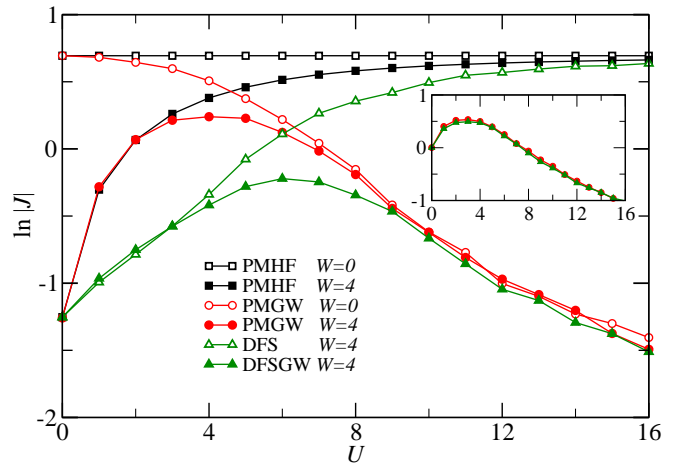}
\caption{As in Fig.~\protect\ref{fig:J_vs_U2D} but for a one dimensional chain with $L=50$ and $W=4t$.  The inset shows PMGW and DFSGW results for a short ($L=8$) chain.}
\label{fig:J_vs_U}
\end{figure}

We begin by showing results for a single configuration of disorder as a function of $U$.
Figures \ref{fig:J_vs_U2D} and \ref{fig:J_vs_U} show the persistent current at half filling as a function of the interaction strength in two and one dimensions respectively.   The disorder strength
is chosen to be equal to the bandwidth of the disorder-free lattice, namely $W=8t$ in two dimensions and $W=4t$ in one dimension.  Results are shown for both the PMGW and DFSGW ground states.  In the two dimensional case, the two approximations produce nearly identical currents, while in one dimension they are quantitatively different, although qualitatively similar.  The distinction  appears to be that $L\ll\xi$ in Fig.~\ref{fig:J_vs_U2D} while $L\gg\xi$ in Fig.~\ref{fig:J_vs_U};  in support of this, we have found that the DFSGW and PMGW currents in one dimension are nearly the same when $L$ is small (inset of Fig.~\ref{fig:J_vs_U}).

The main point of figures~\ref{fig:J_vs_U2D} and \ref{fig:J_vs_U} is that they allow us to distinguish the effects of screening and strong correlations on the persistent current.  
To separate these effects we show, in addition to $J_\mathrm{PMGW}$ and $J_\mathrm{DFSGW}$,  results for calculations without disorder ($W=0$), without strong correlations, and with neither disorder nor strong correlations.  

To suppress strong correlations, the Gutzwiller projection is explicitly turned off. These calculations are labelled ``paramagnetic Hartree-Fock" (PMHF) and ``disordered filled sea" (DFS), and are obtained by setting $g=1$ in the PMGW and DFSGW calculations respectively.   (In the case of the DFS calculations, the screening $\epsilon$ is the same as in the corresponding DFSGW calculations.)     Figures~\ref{fig:J_vs_U2D} and \ref{fig:J_vs_U} show that, as $U$ increases, both $J_\mathrm{PMHF}$ and $J_{DFS}$ increase monotonically, and approach $J$ of the disorder-free system (PMHF with $W=0$).  This indicates that, in the absence of strong correlations, disorder is progressively screened with increasing $U$.      In Fig.~\ref{fig:J_vs_U}, the PMHF curves approach the disorder-free limit faster than the DFS curves, indicating that screening is stronger in the PMHF calculations; however, the qualitative trends are the same in both cases.  

When the Gutzwiller projections are included, the current is nearly identical to the unprojected current for small $U$, and begins to differ from the unprojected current when $U\approx W$.   We have checked several different values of $W$ and found that this is generally the case:  for intermediate disorder strengths, $W$ is the crossover scale beyond which strong correlations begin to play a significant role.  When $U\gtrsim W$, the persistent current is progressively suppressed as $U$ increases.  Within the Gutzwiller approximation, this suppression is interpreted in terms of a quasiparticle spectral weight that is reduced as the Mott transition is approached.  This mechanism is well-known and has been proposed, for example, to explain the anomalously low superfluid screening currents in underdoped cuprate superconductors.\cite{Lee2006}

The most remarkable feature of Figs.~\ref{fig:J_vs_U2D} and \ref{fig:J_vs_U}  is that, although the Gutzwiller projected currents $J_\mathrm{PMGW}$ and $J_\mathrm{DFSGW}$ are suppressed by strong correlations at large $U$, they  approach the  value of the current in the (strongly correlated) disorder-free system.   This indicates that, in spite of the strong correlations, the disorder potential continues to be screened by the interactions. 

We thus have the paradoxical result that the disorder screening {\em increases} with increasing $U$, even though the overall current {\em decreases}.  In contrast, the usual expectation is that disorder screening should become worse as the Mott transition is approached, as occurs for strong disorder. However, our result is consistent with earlier DMFT estimates that suggest that the elastic scattering rate due to weak disorder vanishes at the Mott transition.\cite{Tanaskovic2003}  Our result is also consistent with variational Monte Carlo calculations by Pezzoli et al.,\cite{Pezzoli2010} who used a similar variational approach to study the AHM.  In particular, they
introduced a variational renormalized site energy, similar to the parameter $\epsilon$ used here, and found a monotonic reduction of the effective disorder potential with increasing $U$.

We note that the Gutzwiller calculations shown in Figs.~\ref{fig:J_vs_U2D} and \ref{fig:J_vs_U} break down above some critical interaction strength $U_c(W)$ at which the Mott MIT occurs.
Where this happens depends on the level of disorder, and on the dimension of the lattice.  
For two dimensional clusters with $W=8t$, variational calculations\cite{Pezzoli2010} find a first order Mott transition at  $U_c \sim 12t$, which is well into the regime where both strong correlations and disorder screening are significant.   For  one dimensional systems, the MIT happens at a smaller value; we have performed exact diagonalization calculations for small clusters with $W=4t$ that suggest that $U_c \approx 6t$.   It thus appears that the regime of strong screening and suppressed quasiparticle weight should be most easily observed in two and higher dimensions.

\begin{figure}
\includegraphics[width=\columnwidth]{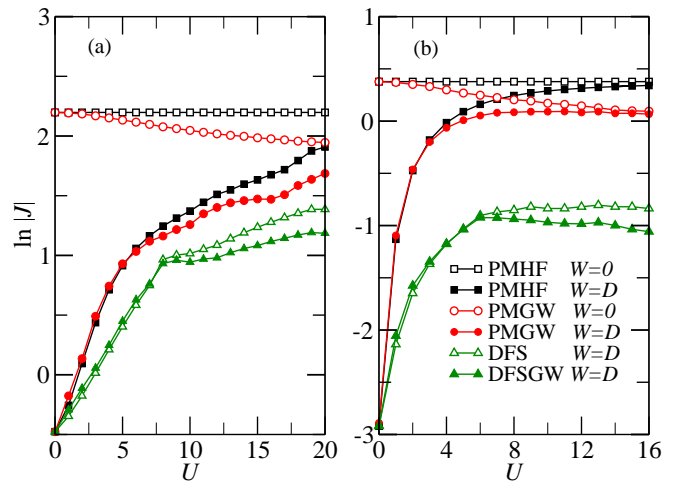}
\caption{As above, but at quarter filling ($n=0.5$).  Results are
for clean ($W=0$) and disordered ($W=D$) systems for (a) a two dimensional lattice with $L=8$ and $D=8t$ and (b) a one dimensional lattice with $L=50$ and $D=4t$.}
\label{fig:quarter_filling}
\end{figure}

For comparison, we show results for the persistent current  at quarter filling in  Fig.~\ref{fig:quarter_filling}.  In this case, there is no Mott transition to worry about, and the Gutzwiller approximation is expected to be meaningful over a large range of $U$ values.\cite{Farhoodfar2009}   The most striking feature of this figure is the diminished role of strong correlations at large $U$, meaning that $J$ is not suppressed significantly at large $U$.

  As with the half-filled case, the figure shows that the PMGW current approaches the disorder-free current in the large $U$ limit.  The DFSGW current, on the other hand, is more weakly screened, and never approaches the clean limit value, but instead decays slowly from a maximum at $U\sim W$ with increasing $U$.   At quarter filling, then, the PMGW and DFSGW approximations are qualitatively different.  In order to determine which result is correct, we have compared the two Gutzwiller results with published exact diagonalization results for the Drude weight at quarter filling in one dimension.\cite{Kotlyar2001}  We find that the $U$-dependence of the DFSGW current in Fig.~\ref{fig:quarter_filling}(b) is surprisingly close to the exact diagonalization results, and in the remainder of this work  we therefore show only DFSGW results at quarter filling.

In summary, Figs.~\ref{fig:J_vs_U2D} and \ref{fig:J_vs_U} illustrate one of the main results of this paper, namely that within the Gutzwiller variational ansatz,  strong correlations do not inhibit the screening of moderate disorder at half filling.

\subsection{Anderson Localization}
As discussed above, we can extract the localization length $\xi$ from the finite size scaling of $J$.  In two dimensions $\xi$ tends to be exponentially large\cite{LeeRMP1985} for weak disorder, and we cannot reach the scaling limit $L>\xi$.    We therefore illustrate our
point by looking at the one dimensional case, where it is possible to obtain $L\gg \xi$, at least for $U$ not too large.  We showed in the previous section that the qualitative roles of screening and correlations are the same in one and two dimensions, and we therefore expect that the main lessons learned about localization in one dimension will also apply to higher dimensions.  To explore the full range of behavior, from weak to strong correlations, we include results for large $U$ where one dimensional systems are actually Mott insulating, but which are accessible in higher dimensions.

\begin{figure}[t]
\includegraphics[width=\columnwidth]{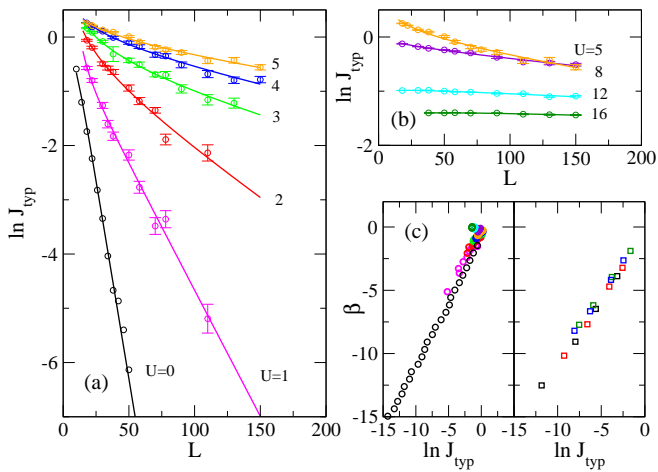}
\caption{Scaling of $\ln J_\mathrm{typ}$ with system size in one dimension.  Typical currents are shown for half filling with (a) $U\lesssim W$ and (b) $U> W$.  Data are for the PMGW approximation, solid lines are fits to the data from Eq.~(\protect\ref{eq:fit}).   Data are calculated for $10000/L$ disorder configurations, and error bars are give the statistical uncertainty in the typical current.  Where error bars are not shown, they are smaller than the symbol size.  (c) The scaling function $\beta$ for PMGW at half filling (left) and  DFSGW at quarter filling (right).  }
\label{fig:scaling}
\end{figure}

Logarithmic plots of the typical current for PMGW calculations at half filling are shown in Fig.~\ref{fig:scaling}.   Lines are fits to the data of the form
\begin{equation}
\ln J_\mathrm{typ} = \ln J_0 - \frac{L}{\xi} + \frac{A}{L^y}
\label{eq:fit}
\end{equation}
where $J_0$, $\xi$, $A$, and $y$ are fitting parameters.  The last term is a finite size correction, meant to account for leading order corrections when $L$ is of order $\xi$.   Clearly, the form of the correction cannot hold for small $L$,  but it improves the quality of the fit significantly over the range of system sizes studied here.

  Two distinct trends can be seen in Fig.~\ref{fig:scaling}: for $U\lesssim W$, the magnitude of $J_\mathrm{typ}$ increases with $U$, but the slope decreases; for $U\gtrsim W$, the magnitude of $J_\mathrm{typ}$ decreases with increasing $U$, and the slope continues to decrease.  Since the slope is the inverse of the localization length, a decreasing slope corresponds to an increasing localization length.  
  
\begin{figure}[t]
\begin{center}
\includegraphics[width=\columnwidth]{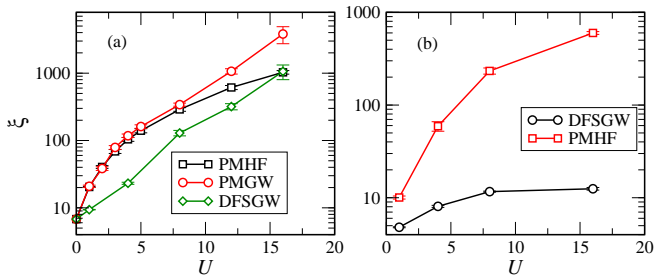}
\end{center}
\caption{Localization length as a function of $U$ for (a) PMGW and DFSGW at half-filling  and (b) DFSGW at quarter-filling.  PMHF results are shown for comparison. }
\label{fig:localizationlength}
\end{figure}

 The localization length is shown in Fig.~\ref{fig:localizationlength}.  At half filling, the PMGW and DFSGW results are qualitatively consistent;  in both cases, the localization length grows monotonically with $U$. 
(Note that where $\xi$ exceeds the largest system sizes, namely $L=150$, the scaling ansatz Eq.~(\ref{eq:fit}) ceases to give quantitatively accurate values for $\xi$, and the results are qualitative only.) 
  There is thus a region of the phase diagram where the overall magnitude of the current is suppressed by strong correlations, but the localization length grows with $U$.  In contrast, at quarter filling, the localization length saturates at large $U$.

In Fig.~\ref{fig:scaling}(c), we plot the scaling function
\begin{equation}
\beta \equiv \frac{d\ln J_\mathrm{typ}}{d \ln L}
\label{eq:scaling}
\end{equation}
as a function of $\ln J_\mathrm{typ}$ for both half filling and quarter filling.  Note that the derivative in Eq.~(\ref{eq:scaling}) is obtained from the fitted curves in Figs.~\ref{fig:scaling} (a) and (b). 
When one-parameter scaling holds, $\beta$ is a single-valued function of $J_\mathrm{typ}$.   The main branch of the data shown in Fig.~\ref{fig:scaling}(c) corresponds to $U<W$ and is, within the accuracy of our data, a function of $J_\mathrm{typ}$ alone.  There is also a small hook-shaped feature in the upper right hand corner of the figure that corresponds to a second branch which comes from $U>W$.  This branch bends back towards the left, making $\beta$ a multi-valued function of $J_\mathrm{typ}$.   At half filling, it appears that one parameter scaling breaks down when $U>W$.  In contrast, the quarter-filling data lies entirely on the main branch, and therefore appears to satisfy one-parameter scaling.

This breakdown of one parameter scaling at half filling follows directly from the two competing tendencies illustrated in Figs.~\ref{fig:J_vs_U2D} and \ref{fig:J_vs_U}.  The main branch of the $\beta$ function comes from $U<W$, where $J$ depends on $U$ primarily through disorder screening,  while the second branch comes from $U>W$ where strong correlations are significant.  While this breakdown may be hard to observe in one dimension because of the relatively low value of $U_c$, it should be observable in two and higher dimensions.

We note that our DFSGW results for $\xi$ at quarter filling are entirely consistent with published DMRG calculations by Nishimoto et al.\cite{Nishimoto2010}    It is difficult to compare results at half-filling, however, because the authors perform their DMRG calculations for $U> U_c(W)$, and calculate $\xi$ as a function of filling, while enhanced disorder screening is predicted to occur at half-filling as $U\rightarrow U_c$.  

\section{Conclusions}
\label{Sec:conclusions}

In summary, we have studied the effects of disorder and strong correlations on transport in the Anderson Hubbard model.  We have focused on the interaction-driven Mott transition at half filling with weak and intermediate strength disorder.  
We used a Gutzwiller variational ansatz to describe the strongly correlated state near the Mott transition, and studied persistent currents induced by an Aharonov-Bohm flux.  

We found that the persistent current is a nonmonotonic function of $U$.  For small $U$, the Coulomb interactions screen the disorder potential and the current is an increasing function of $U$.  For large $U$, strong correlations suppress the current, which becomes a decreasing function of $U$.  The crossover between these two limits occurs at $U\approx W$.  The main surprise in our calculations is that disorder screening persists when $U$ is large.  One consequence of this is that in the strongly correlated regime ($U>W$) the localization length is an increasing function of $U$, even though the current itself is a decreasing function of $U$.  We expect that this apparently-paradoxical behavior should be observable two and higher dimensions.

\section*{Acknowledgments}
RJG  and WAA acknowledge NSERC of Canada for their support.   This work was made possible by the facilities of the Shared Hierarchical Academic Research Computing Network (SHARCNET:www.sharcnet.ca) and Compute/Calcul Canada.


\end{document}